# Penalized maximum likelihood for multivariate Gaussian mixture

Hichem Snoussi* and Ali Mohammad-Djafari*

*Laboratoire des Signaux et Systèmes (L2S),
Supélec, Plateau de Moulon, 91192 Gif-sur-Yvette Cedex, France

**Abstract.** In this paper, we first consider the parameter estimation of a multivariate random process distribution using multivariate Gaussian mixture law. The labels of the mixture are allowed to have a general probability law which gives the possibility to modelize a temporal structure of the process under study. We generalize the case of univariate Gaussian mixture in [1] to show that the likelihood is unbounded and goes to infinity when one of the covariance matrices approaches the boundary of singularity of the non negative definite matrices set. We characterize the parameter set of these singularities. As a solution to this degeneracy problem, we show that the penalization of the likelihood by an Inverse Wishart prior on covariance matrices results to a penalized or maximum *a posteriori* criterion which is bounded. Then, the existence of positive definite matrices optimizing this criterion can be guaranteed. We also show that with a modified EM procedure or with a Bayesian sampling scheme, we can constrain covariance matrices to belong to a particular subclass of covariance matrices. Finally, we study degeneracies in the source separation problem where the characterization of parameter singularity set is more complex. We show, however, that Inverse Wishart prior on covariance matrices eliminates the degeneracies in this case too.

## INTRODUCTION

We consider a double stochastic process:

- A discrete process $(z_t)_{t=1..T}$, with $z_t$ taking its values in the discrete set $\mathcal{Z} = \{1..K\}$.
- A continuous process $(\boldsymbol{s}_t)_{t=1..T}$ which is white conditionally to the first process $(z_t)_{t=1..T}$ and following a distribution:

  $$p(\boldsymbol{s}|z) = f(\boldsymbol{s}; \zeta_z)$$

  In the following, without loss of generality of the considered model, we restrict the function $f(.)$ to be a Gaussian: $f(.|z) = \mathcal{N}(\boldsymbol{\mu}_z, \boldsymbol{R}_z)$.

This double process is called in literature "Mixture model". When the hidden process $z_{1..T}$ is white, we have an i.i.d mixture model: $p(\boldsymbol{s}) = \sum_z p_z \mathcal{N}(\boldsymbol{\mu}_z, \boldsymbol{R}_z)$ and when $z_{1..T}$ is Markovian, the model is called HMM (Hidden Markov Model). For application of these two models see [2] and [3]. Mixture models present an interesting alternative to non parametric modeling. By increasing the number of mixture components, we are able to approximate any probability density and the time dependence structure of the hidden process $z_{1..T}$ allows to take into account a correlation structure of the resulting process. In the following, for clarity of demonstrations, we assume an i.i.d. mixture model.

## CHARACTERIZATION OF LIKELIHOOD DEGENERACY

We consider $T$ observations $(\boldsymbol{s}_t)_{t=1..T}$ of a random $n$-vector following a multivariate Gaussian mixture law:

$$p(\boldsymbol{s}_t) = \sum_{z=1}^{K} p_z \mathcal{N}(\boldsymbol{s}_t; \boldsymbol{\mu}_z, \boldsymbol{R}_z)$$

Where $p_z = P(Z = z)$ is the probability that the random hidden label $Z$ takes the value $z \in \mathcal{Z} = \{1..K\}$, $\boldsymbol{\mu}_z$ is the $n$-vector mean of the Gaussian component $z$ and $\boldsymbol{R}_z$ its $n \times n$ covariance matrix. We intend to estimate the parameters $\boldsymbol{\theta}_z = (p_z, \boldsymbol{\mu}_z, \boldsymbol{R}_z)_{z \in 1..K}$ by maximizing its likelihood given the observations $\boldsymbol{s}_{1..T} = [\boldsymbol{s}_t]_{t=1..T}$:

$$\widehat{\boldsymbol{\theta}} = \arg\max_{\boldsymbol{\theta} \in \boldsymbol{\Theta}} p(\boldsymbol{s}_{1..T} \,|\, \boldsymbol{\theta})$$

Where

$$p(\boldsymbol{s}_{1..T} \,|\, \boldsymbol{\theta}) = \prod_{t=1}^{T} \sum_{z} p_z \, |2\pi \boldsymbol{R}_z|^{(-1/2)} \exp\left[ -\frac{1}{2} (\boldsymbol{s}_t - \boldsymbol{\mu}_z)^T \boldsymbol{R}_z^{-1} (\boldsymbol{s}_t - \boldsymbol{\mu}_z) \right]$$

and

$$\boldsymbol{\Theta} = \left\{ \boldsymbol{\theta}_z = (p_z, \boldsymbol{\mu}_z, \boldsymbol{R}_z) \,|\, p_z \in \mathrm{R}_+, \textstyle\sum_{z=1}^{K} p_z = 1; \boldsymbol{R}_z \in \mathcal{R}; \boldsymbol{\mu}_z \in \mathrm{R}^n \right\} \tag{1}$$

$\mathcal{R}$ is a closed subset of covariance matrices. Some examples of $\mathcal{R}$ are considered later in section $4$ and in [4].

**Proposition** $1$ [Likelihood function is unbounded]: $\forall\ \boldsymbol{s}_{1..T} \in (\mathrm{R}^n)^T$, $\exists$ a singularity point $\boldsymbol{\theta}_s$ in the parameter space $\boldsymbol{\Theta}$ such that: $\lim_{\boldsymbol{\theta} \to \boldsymbol{\theta}_s} p(\boldsymbol{s}_{1..T} \,|\, \boldsymbol{\theta}) = \infty$. These points are the $\boldsymbol{\theta} = (p_z, \boldsymbol{\mu}_z, \boldsymbol{R}_z)_{\boldsymbol{z} \in \mathcal{Z}}$ such that, at least one of the $\boldsymbol{R}_z$ (but not all of them together) is a singular non negative matrix and the correspondent mean $\boldsymbol{\mu}_z$ lies in the intersection of $n - rank(\boldsymbol{R}_z)$ hyperplans of $\mathrm{R}^n$.

**Proof**: Let $z_0 \in \mathcal{Z}$ and $\boldsymbol{R}_{z_0}$ be a singular NND matrix of rank $p < n$. $\boldsymbol{R}_{z_0}$ can be diagonalized in the orthogonal group:

$$\boldsymbol{R}_{z_0} = \boldsymbol{U}^T \boldsymbol{\Lambda} \boldsymbol{U}, \qquad \boldsymbol{\Lambda} = \begin{bmatrix} 0 & & & & & \\ & \ddots & & & & \\ & & 0 & & & \\ & & & \lambda_{n-p+1} & & \\ & & & & \ddots & \\ & & & & & \lambda_n \end{bmatrix}$$

Consider now a sequence of positive definite matrices $\left(\boldsymbol{R}_{z_0}^{(n)}\right)_{n\in\mathrm{N}}$ defined by:

$$\boldsymbol{R}_{z_0}^{(n)} = \boldsymbol{U}^T \begin{bmatrix} \lambda_1^{(n)} & & & & & \\ & \ddots & & & & \\ & & \lambda_{n-p}^{(n)} & & & \\ & & & \lambda_{n-p+1} & & \\ & & & & \ddots & \\ & & & & & \lambda_n \end{bmatrix} \boldsymbol{U}$$

With the $(n-p)$ strictly positive numeric sequences $\left(\lambda_i^{(n)}\right)_{i=1..(n-p)}$ which tend to $0$. Thus the sequence of $\left(\boldsymbol{R}_{z_0}^{(n)}\right)_{n\in\mathrm{N}}$ converges to $\boldsymbol{R}_{z_0}$. Likelihood function evaluated at $\boldsymbol{R}_{z_0}^{(n)}$ is:

$$\begin{aligned} p_n(\boldsymbol{s}_{1..T}\,|\,\boldsymbol{\theta}) = \prod_{t=1}^{T} \Bigg( & p_{z_0}\,|2\pi\boldsymbol{R}_{z_0}^{(n)}|^{(-1/2)} \exp\left[-\frac{1}{2}(\boldsymbol{s}_t-\boldsymbol{\mu}_{z_0})^T {\boldsymbol{R}_{z_0}^{(n)}}^{-1}(\boldsymbol{s}_t-\boldsymbol{\mu}_{z_0})\right] \\ & + \sum_z p_z \mathcal{N}(\boldsymbol{\mu}_z, \boldsymbol{R}_z)\Bigg) \end{aligned}$$

Expending the exponent of the component $z_0$ in canonical form :

$$(\boldsymbol{s}_t-\boldsymbol{\mu}_z)^T {\boldsymbol{R}_{z_0}^{(n)}}^{-1}(\boldsymbol{s}_t-\boldsymbol{\mu}_z) = \sum_i \frac{[\boldsymbol{U}(\boldsymbol{s}_t-\boldsymbol{\mu}_z)]_i^2}{\lambda_i^{(n)}},$$

We can see that when the eigenvalues $(\lambda_i^{(n)})_{i=1..(n-p)}$ tend to zero, or equivalently, when the covariance $\boldsymbol{R}_{z_0}^{(n)}$ tends to $\boldsymbol{R}_{z_0}$ and when $\boldsymbol{\mu}_{z_0}$ lies in the intersection of the hyperplans $(\mathcal{H}_i = \{\boldsymbol{\mu}\,|\,[\boldsymbol{U}(\boldsymbol{s}_t-\boldsymbol{\mu})]_i = 0\})_{i=1..(n-p)}$, the likelihood function goes to infinity. So we have proved that any singular NND matrix is a point of degeneracy provided that the means lie in specific hyperplans. In one dimensional case, this corresponds to the fact that $\sigma$ goes to zero and the correspondent mean coincides with one observation.

Figure $1$ shows an example of this degeneracy. In this example, we take an original distribution of a $2$-D random vector which is a mixture of $10$ Gaussians. The Gaussians have their means located on a cercle and have the same covariance. Figure $1$-a shows the graph of this distribution from which we generated $100$ samples in order to estimate its parameters. Figure $1$-b shows the estimated distribution. We can note the failure of the maximum likelihood estimator and its tendency to converge to sharp Gaussians.

Here, we highlight the effect of growing the dimension $n$ which increases the occurrence of degeneracy. We have, for $n > 1$ an infinite number of singularities. Moreover, even if we fix the means of the mixture components, the unboundedness of likelihood might occur if some covariances go to particular singular matrices . But, we think that this second kind of degeneracy is less likely to happen particularly if the number of samples grows. We note that the occurrence of degeneracy increases when the dimension grows and decreases when the number of samples grows.

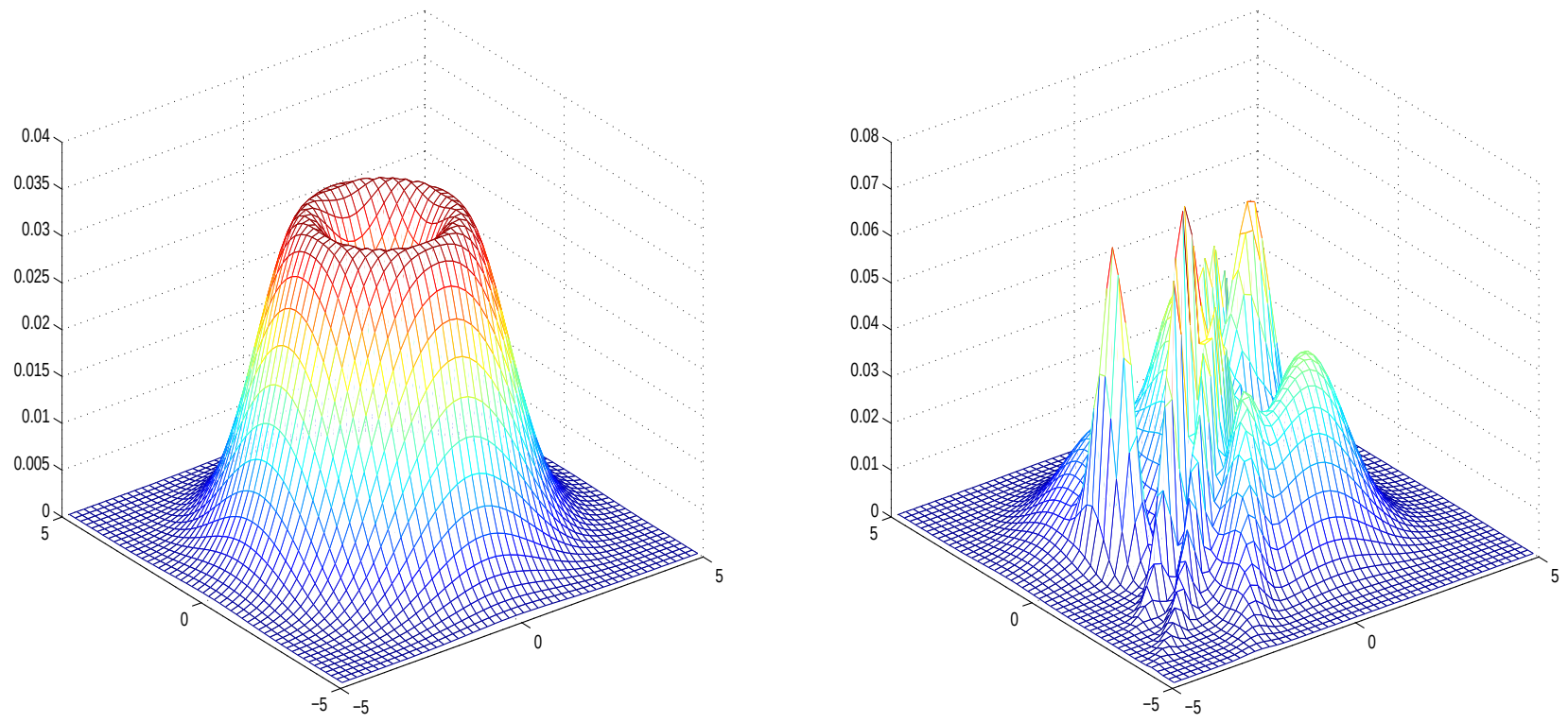

Original distribution

ML Estimated distribution with $100$ samples

Fig-$1$. Failure of the ML estimation of the parameters of a $10$ component Gaussian mixture distribution.

## BAYESIAN SOLUTION TO DEGENERACY

This degeneracy was noted by many authors (Day, 1969 [5]) and in (Hathaway 1986 [6]), a constraint formulation of the EM algorithm has been proposed to eliminate this degeneracy. In (Ormoneit 1998 [7]), a penalization by an Inverse Wishart prior was employed to eliminate it. Our contribution leads to the same penalization but in different manner. In (Ormoneit 1998), the Inverse Wishart prior was chosen because it is a conjugate prior. In the one dimensional case [1], the penalization by an Inverse Gamma prior on variances was used to eliminate degeneracy.

In this work, after characterizing the origin of these singularities, we extend this procedure to the multivariate case to propose an Inverse Wishart prior on covariances $\boldsymbol{R}_z$ which guarantees the boundness of the likelihood:

$$p_{\alpha,\beta,\boldsymbol{J}}(\boldsymbol{R}_z) = \frac{K}{|\boldsymbol{R}_z|^{\beta}} \exp\left[-\alpha\,\mathrm{Tr}\left(\boldsymbol{R}_z{}^{-1}\boldsymbol{J}\right)\right]$$

where K is a normalization constant, $\alpha$ and $\beta$ two strictly positive constants which contain *a priori* information about the power level (scale parameter) and $\boldsymbol{J}$ is a positive definite symmetric matrix which contains *a priori* information on the covariance structure. In fact, the mode of this law is given by:

$$\frac{\partial \log\left[p(\boldsymbol{R}_z)\right]}{\partial \boldsymbol{R}_z} = -\beta \boldsymbol{R}_z^{-1} + \alpha \boldsymbol{R}_z^{-1} \boldsymbol{J} \boldsymbol{R}_z^{-1} = 0$$

Leading to:

$$\boldsymbol{R}_z = \frac{\alpha}{\beta}\boldsymbol{J}$$

**Proposition** 2**:** $\forall\, \boldsymbol{s}_{1..T} \in (\mathrm{R}^n)^T$, the *a posteriori* distribution $p(\boldsymbol{\theta} \,|\, \boldsymbol{s}_{1..T})$ with the *a priori*:

$$p(\boldsymbol{\theta}) = \prod_{\boldsymbol{z} \in \mathcal{Z}} p_{\alpha_z, \beta_z, \boldsymbol{J}_z}(\boldsymbol{R}_z)$$

is bounded and goes to $0$ when one of the covariance matrices $\boldsymbol{R}_z$ approaches a singular matrix.

**Proof:** The penalized likelihood is:

$$p(\boldsymbol{s}_{1..T} \,|\, \boldsymbol{\theta})\, p(\boldsymbol{\theta}) = \prod_{t=1}^{T} \left( (\prod_{\boldsymbol{z} \in \mathcal{Z}} p(\boldsymbol{R}_z))^{1/T} \sum_{z} p_z \mathcal{N}(\boldsymbol{\mu}_z, \boldsymbol{R}_z) \right)$$

For each label $z$, we have the following inequality:

$$(\prod_{\boldsymbol{z} \in \mathcal{Z}} p(\boldsymbol{R}_z))^{1/T} \mathcal{N}(\boldsymbol{\mu}_z, \boldsymbol{R}_z) \leq \frac{A}{|\boldsymbol{R}_z|^{1/2}} \prod_{\boldsymbol{z} \in \mathcal{Z}} \frac{K_z}{|\boldsymbol{R}_z|^{\beta_z}} \exp\left[-\alpha_z \,\mathrm{Tr}\left(\boldsymbol{R}_z^{-1} \boldsymbol{J}_z\right)\right]$$

Thus, to prove the proposition, we need to show that $\forall\, a > 0$, $b > 0$ and $\boldsymbol{R}_s$ a singular matrix, we have:

$$\lim_{\boldsymbol{R}_z \longrightarrow \boldsymbol{R}_s} \frac{1}{|\boldsymbol{R}_z|^b} \exp\left[-a \,\mathrm{Tr}\left(\boldsymbol{R}_z^{-1} \boldsymbol{J}\right)\right] = 0$$

Using the inequality

$$(det\, \boldsymbol{A})^{1/n} \leq \frac{1}{n} \mathrm{Tr}\,(\boldsymbol{A})$$

valid for any real symmetric $n \times n$ matrix $\boldsymbol{A}$, We have:

$$\frac{1}{|\boldsymbol{R}_z|^b} \exp\left[-a \,\mathrm{Tr}\left(\boldsymbol{R}_z^{-1} \boldsymbol{J}\right)\right] \leq \frac{1}{|\boldsymbol{R}_z|^b} \exp\left[-a\, n \frac{|\boldsymbol{J}|^{1/n}}{|\boldsymbol{R}_z|^{1/n}}\right]$$

In the above inequality, the right hand side term goes to zero when $\boldsymbol{R}_z$ approaches the boundary of singularity. Therefore, the penalized likelihood is bounded and is null on the boundary of singularity.

At this point, we can also follow the arguments in [4] to prove the existence of positive definite matrices corrresponding to the modes of the penalized likelihood. Figure 2 illustrates the regularization effect of this penalization. Here we used the same samples generated for the figure 1 and estimated the parameters of the mixture by optimizing the penalized likelihood criterion. The probability of degeneracy is zero.

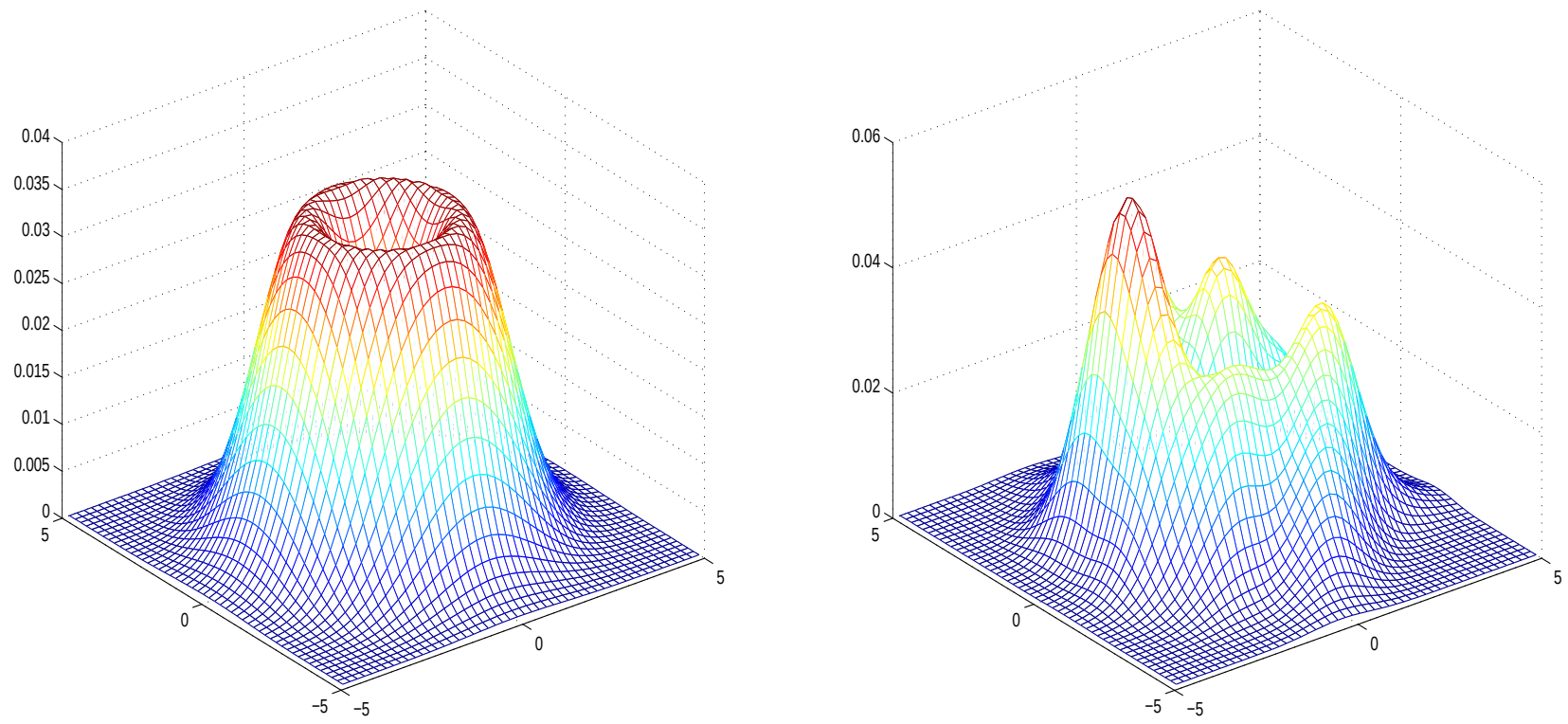

Original distribution

Penalized EM Estimated distribution with $100$ samples

Fig-$2$. Regularization effect of the penalized EM algorithm.

## ESTIMATION OF STRUCTURED COVARIANCE MATRICES

In this paragraph, we generalize the work in [4] to estimate covariance matrices of specified structure in the mixture case. The constraints are summarized in the closed subset $\mathcal{R}$ introduced in the definition of the parameter set $\boldsymbol{\Theta}$ (1).

### Unconstrained case:

The unconstrained case was treated in many works. In [7], three methods were proposed: Averaging, maximum penalized likelihood and Bayesian sampling. We briefly recall the EM algorithm and the Bayesian sampling which both can be seen as data augmentation algorithms:

- EM algorithm: It consists of two steps:
  **(i)** E (Expectation)-step: Consider observations $\boldsymbol{s}_{1..T}$ as incomplete data and $(\boldsymbol{s}_{1..T}, z_{1..T})$ as complete data and compute the functional $\mathcal{Q}(\boldsymbol{\theta} \,|\, \boldsymbol{\theta}^{(k)}) = E\{\log p(\boldsymbol{s}_{1..T}, z_{1..T} \,|\, \boldsymbol{\theta}) + \log p(\boldsymbol{\theta}) \,|\, \boldsymbol{s}_{1..T}, \boldsymbol{\theta}^{(k)}\}$;
  **(ii)** M (Maximization)-step: Update $\boldsymbol{\theta}^{(k+1)} = \arg\max_{\boldsymbol{\theta}} \mathcal{Q}(\boldsymbol{\theta} \,|\, \boldsymbol{\theta}^{(k)})$.

- Bayesian sampling: It consists of two steps:
  **(i)** Generate $z_{1..T}^{(k+1)} \sim p(z_{1..T} \,|\, \boldsymbol{s}_{1..T}, \boldsymbol{\theta}^{(k)})$;
  **(ii)** Generate $\boldsymbol{\theta}^{k+1} \sim p(\boldsymbol{\theta} \,|\, \boldsymbol{s}_{1..T}, z_{1..T}^{(k+1)})$.

In the unconstrained case, one obtains, in both first steps of the above algorithms, functionals which have only one maximum obtained by canceling the gradient to zero.

## Constrained case:

In both EM algorithm and Bayesian sampling methods presented above, the second step which consists in updating $\boldsymbol{\theta}$ was unconstrained. We see in the following how we are able to combine the data augmentation algorithms with the iterative gradient algorithm proposed in [4] to constrain the covariance matrix $\boldsymbol{R}_z$ to be in the closed set $\mathcal{R}$.

### *Strutured EM*

The functional $\mathcal{Q}(\boldsymbol{\theta} \,|\, \boldsymbol{\theta}^{(k)})$ can be decomposed as follows:

$$\mathcal{Q}(\boldsymbol{\theta} \,|\, \boldsymbol{\theta}^{(k)}) = \sum_{z=1}^{K} g(\boldsymbol{R}_z, \boldsymbol{S}_z) + f(\boldsymbol{p}, \boldsymbol{\mu} \,|\, \boldsymbol{\theta}^{(k)})$$

with:

$$\begin{cases} g(\boldsymbol{R}_z, \boldsymbol{S}_z) = -(1 + \frac{\beta}{N_z}) \log |\boldsymbol{R}_z| - \mathrm{Tr}\left( \boldsymbol{R}_z^{-1} \left( \boldsymbol{S}_z + \frac{\alpha \boldsymbol{J}}{N_z} \right) \right) \\ N_z = \sum_{t=1}^{T} p(z(t) = z \,|\, \boldsymbol{s}(t), \boldsymbol{\theta}^{(k)}) \end{cases}$$

and $\boldsymbol{S}_z$ the weighted sample covariance matrix:

$$\boldsymbol{S}_z = \frac{\sum_{t=1}^{T} (\boldsymbol{s}(t) - \boldsymbol{\mu}_z^{(k+1)}) \, (\boldsymbol{s}(t) - \boldsymbol{\mu}_z^{(k+1)})^* \, p(z(t) = z \,|\, \boldsymbol{s}(t), \boldsymbol{\theta}^{(k)})}{\sum_{t=1}^{T} p(z(t) = z \,|\, \boldsymbol{s}(t), \boldsymbol{\theta}^{(k)})}$$

Thus, the maximization of $\mathcal{Q}$ with respect to $\boldsymbol{R}_z$ is equivalent to the maximization of $g(\boldsymbol{R}_z, \boldsymbol{S}_z)$ with respect to $\boldsymbol{R}_z$. The necessary gradient equations are:

$$\delta g(\boldsymbol{R}_z, \boldsymbol{S}_z) = \mathrm{Tr}\left( (\boldsymbol{R}_z^{-1} (\boldsymbol{S}_z + \frac{\alpha \boldsymbol{J}}{N_z}) \boldsymbol{R}_z^{-1} - (1 + \frac{\beta}{N_z}) \boldsymbol{R}_z^{-1}) \delta \boldsymbol{R}_z \right) = 0 \qquad (2)$$

In the unconstrained case, the solution of (2) is $\boldsymbol{R}_z = \frac{\boldsymbol{S}_z + \frac{\alpha \boldsymbol{J}}{N_z}}{1 + \frac{\beta}{N_z}}$. Constraint maximization of $g$ with $\boldsymbol{R}_z \in \mathcal{R}$ for any $\mathcal{R}$ is not easy. However, if $\mathcal{R}$ is such that $\boldsymbol{R} \in \mathcal{R} \Rightarrow \delta \boldsymbol{R} \in \mathcal{R}$ (for example the set of Toeplitz matrices) then we replace the second step of the EM algorithm by the following:

1. Find $\boldsymbol{D}_z^{(k+1)}$ belonging to $\mathcal{R}$ so that $g(\boldsymbol{R}_z^{(k)}, \boldsymbol{S}_z - \boldsymbol{D}_z^{(k+1)})$ satisfies the necessary gradient conditions.
2. Put $\boldsymbol{R}_z^{(k+1)} = \boldsymbol{R}_z^{(k)} + \boldsymbol{D}_z^{(k+1)}$

This modification preserves the monotonicity of the EM algorithm and makes the problem linear in $\boldsymbol{D}_z$ and so it is easier to impose constraints with the condition that the variation of $\boldsymbol{R}_z$ still belongs to $\mathcal{R}$, which is true for a wide range of constraints such in the Toeplitz case.

*Structured Bayesian sampling*

We propose the following Bayesian sampling scheme:

1. Generate $z^*_{1..T} \sim p(z_{1..T} \,|\, \boldsymbol{s}_{1..T}, \boldsymbol{\theta}^{(k)})$;
2. Generate $\boldsymbol{D}_z^{(k+1)}$ belonging to $\mathcal{R}$ according to the *a posteriori* distribution $p(\boldsymbol{D}_z \,|\, \boldsymbol{s}_{1..T}, z^*_{1..T}) \sim \exp\left[g(\boldsymbol{R}_z^{(k)}, \boldsymbol{S}_z - \boldsymbol{D}_z^{(k+1)})\right]$.
3. Update $\boldsymbol{R}_z^{(k+1)} = \boldsymbol{R}_z^{(k)} + \boldsymbol{D}_z^{(k+1)}$

$\boldsymbol{S}_z$ is the sample covariance depending on the partition defined by $z^*_{1..T}$:

$$\begin{cases} \boldsymbol{S}_z = \frac{\sum_{t\in\mathcal{T}_z} \boldsymbol{s}(t)\boldsymbol{s}(t)^*}{Card(\mathcal{T}_z)} \\ \mathcal{T}_z = \{t \,|\, z(t) = z\} \end{cases}$$

To be sure that the sampling keeps $\boldsymbol{D}_z$ in the closed set $\mathcal{R}$, we define a basis $(\boldsymbol{Q}_l)_{l=1..L}$ of $\mathcal{R}$ and we sample the projection of $\boldsymbol{D}_z$ on $\mathcal{R}$: $\boldsymbol{x}_{1..L} \sim p(\boldsymbol{x}_{1..L} \,|\, \boldsymbol{s}_{1..T}, z^*_{1..T})$, where the vector $\boldsymbol{x}_{1..L}$ is defined as:

$$\boldsymbol{D}_z = \sum_{l=1}^{L} x_l \, \boldsymbol{Q}_l$$

## MIXED SOURCES

We consider now the case where sources are not directly observed, but mixed with an unknown mixing matrix $\boldsymbol{A}$ and we want to take into account measurement errors so that observations are modeled by the following equation:

$$\boldsymbol{x}(t) = \boldsymbol{A}\,\boldsymbol{s}(t) + \boldsymbol{n}(t)$$

In this section, we show that when we are interested in estimating jointly the mixing matrix $\boldsymbol{A}$, noise covariance matrix $\boldsymbol{R}_\epsilon$ and the parameters of the mixture, by maximizing the likelihood $p(\boldsymbol{x}_{1..T} \,|\, \boldsymbol{A}, \boldsymbol{R}_\epsilon, \boldsymbol{\theta}_z)$, we encounter the same problems of degeneracy mentioned above. Likelihood function has the following expression:

$$p(\boldsymbol{x}_{1..T} \,|\, \boldsymbol{A}, \boldsymbol{R}_\epsilon, \boldsymbol{\theta}_z) = \prod_{t=1}^{T} \sum_{z=1}^{K} p_z(t) \mathcal{N}(\boldsymbol{A}\boldsymbol{\mu}_z, \, \boldsymbol{A}\boldsymbol{R}_z\boldsymbol{A}^* + \boldsymbol{R}_\epsilon)$$

with $\boldsymbol{\theta}_z = (\boldsymbol{\mu}_z, \boldsymbol{R}_z, p_z)$.

The expression $p_z(t) = \sum\limits_{z_{1..T}, z(t)=z} p(z_{1..T})$ represents the marginal law of $z(t)$. Indeed, the hidden variables do not need necessarily to be white and so the mixture to be i.i.d. We can rewrite the expression of the likelihood in a more general form in which the marginalization is not performed :

$$p(\boldsymbol{x}_{1..T}\,|\,\boldsymbol{A},\boldsymbol{R}_\epsilon,\boldsymbol{\theta}_z)=\sum_{z_{1..T}}p(z_{1..T})\prod_{t=1}^{T}\mathcal{N}(\boldsymbol{A}\boldsymbol{\mu}_z,\,\boldsymbol{A}\boldsymbol{R}_z\boldsymbol{A}^*+\boldsymbol{R}_\epsilon)$$

It is obvious, under this form, that degeneracy happens when one of the terms constituting the sum tends to infinity and this is independently of the law $p(z_{1..T})$.

Consider now the matrices $\boldsymbol{\Gamma}_z=\boldsymbol{A}\boldsymbol{R}_z\boldsymbol{A}^*+\boldsymbol{R}_\epsilon$. It's clear that degeneracy is produced when, among matrices $\boldsymbol{\Gamma}_z$, at least one is singular and one is regular. We show in the following that this situation can occur.

We recall that the matrices $\boldsymbol{R}_z$ and $\boldsymbol{R}_\epsilon$ belong to a closed subset of the set of the non negative definite matrices. Constraining matrices to be positive definite leads to complicated solutions. The main origin of this complication is the fact that the set of positive definite matrices is not closed. For the same reason, we don't constrain the mixing matrix $\boldsymbol{A}$ to be of full rank.

**Proposition 3**: $\forall\,\boldsymbol{A}$ non null, $\exists$ matrices $\{\boldsymbol{\Gamma}_z=\boldsymbol{A}\boldsymbol{R}_z\boldsymbol{A}^*+\boldsymbol{R}_\epsilon$ for $z=1..K\}$ such that $\{z\mid\boldsymbol{\Gamma}_z\;is\;singular\}\neq\emptyset$ and $\{z\mid\boldsymbol{\Gamma}_z\;is\;regular\}\neq\emptyset$.
$\boldsymbol{R}_\epsilon$ is necessarily a singular NND matrix and $Card\,(\{z\mid\boldsymbol{R}_z\;is\;regular\})<K$.

**Proof**: Without affecting the generality of the problem, we show how to construct a singular matrix $\boldsymbol{\Gamma}_1$ and the others matrices $\boldsymbol{\Gamma}_z$ regular. We consider NND matrices. Therefore, the kernel of the correspondent linear mapping coincides with its isotropic cone. Thus, we have:

$$Ker(\boldsymbol{\Gamma}_z)=Ker(\boldsymbol{A}\boldsymbol{R}_z\boldsymbol{A}^*)\cap Ker(\boldsymbol{R}_\epsilon)$$

It is sufficient to prove the existence of $\boldsymbol{R}_\epsilon$ and $(\boldsymbol{R}_z)_{z=1..K}$ that verify the following condition:

$$\left\{\begin{array}{lcl}Ker(\boldsymbol{A}\boldsymbol{R}_1\boldsymbol{A}^*)\cap Ker(\boldsymbol{R}_\epsilon) & \neq & \{0\}\\ Ker(\boldsymbol{A}\boldsymbol{R}_z\boldsymbol{A}^*)\cap Ker(\boldsymbol{R}_\epsilon) & = & \{0\}\,,\,\boldsymbol{z}=2..K\end{array}\right.\tag{3}$$

If the matrix $\boldsymbol{R}_\epsilon$ is regular, there is no degeneracy: According to the mini-max principle applied to the characterization of the eigenvalues of the sum of two hermitian matrices, the eigenvalues of $\boldsymbol{\Gamma}_z$ are greater than those of $\boldsymbol{R}_\epsilon$ and then strictly positive which imply that all of the matrices $\boldsymbol{\Gamma}_z$ are regular.

We have:

$$Ker(\boldsymbol{A}^*)\subseteq Ker(\boldsymbol{A}\boldsymbol{R}_z\boldsymbol{A}^*),\;z=1..K\tag{4}$$

Equality holds if $\boldsymbol{R}_z$ is regular or if $Ker(\boldsymbol{R}_z)\cap Im(\boldsymbol{A}^*)=\{0\}$. Note that if all the matrices $\boldsymbol{R}_z$ are regular, there is no degeneracy.

Suppose then that the matrices $\boldsymbol{R}_z$, except the first matrix $\boldsymbol{R}_1$, are regular. We will try now to construct the matrices $\boldsymbol{R}_1$ and $\boldsymbol{R}_\epsilon$ making the condition (3) verified. Let a non null vector $\boldsymbol{x}_s$ belong to $[Ker(\boldsymbol{A}^*)]^{\perp}$. There exist NND matrices $\boldsymbol{R}_1$ and $\boldsymbol{R}_\epsilon$ such that $\boldsymbol{x}_s \in Ker(\boldsymbol{A}\boldsymbol{R}_1\boldsymbol{A}^*) \cap Ker(\boldsymbol{R}_\epsilon)$. In fact, consider the family of vectors $(\boldsymbol{x}_j)_{j\in J}$ belonging to $Ker(\boldsymbol{A}^*)$ such that the family $\{\boldsymbol{x}_s\} \cup (\boldsymbol{x}_j)_{j\in J}$ is orthogonal (this is insured by the principle of the incomplete basis). The matrices $\boldsymbol{R}_1 = \sum_{j\in J} \alpha_j \boldsymbol{x}_j \boldsymbol{x}_j^*$ ($\alpha_j \geq 0$) and $\boldsymbol{R}_\epsilon = \sum_{j\in J} \beta_j \boldsymbol{x}_j \boldsymbol{x}_j^*$ ($\beta_j \geq 0$) are such that $\boldsymbol{x}_s \in Ker(\boldsymbol{A}\boldsymbol{R}_1\boldsymbol{A}^*) \cap Ker(\boldsymbol{R}_\epsilon)$ by construction and $Ker(\boldsymbol{A}\boldsymbol{R}_z\boldsymbol{A}^*) \cap Ker(\boldsymbol{R}_\epsilon) = \{0\}$. We have then constructed matrices which verify the degeneracy condition.

Note that the fact that the matrices $\boldsymbol{R}_1$ and $\boldsymbol{R}_\epsilon$ are singular is a necessary condition but not sufficient; the matrix $\boldsymbol{R}_1$ can be singular with $Ker(\boldsymbol{A}\boldsymbol{R}_1\boldsymbol{A}^*) = Ker(\boldsymbol{A}^*)$ and so there is no degeneracy, or as well, $\boldsymbol{R}_\epsilon$ is singular but $Ker(\boldsymbol{A}\boldsymbol{R}_z\boldsymbol{A}^*) \cap Ker(\boldsymbol{R}_\epsilon) \neq \{0\}$, $\forall z \in \{1..K\}$, which implies that all matrices $\boldsymbol{\Gamma}_z$ are singular and so no degeneracy occurs.

## DEGENERACY ELIMINATION IN THE MIXED CASE

In the light of what we presented in the two first paragraphs, one possible way to eliminate this degeneracy consists in penalizing the likelihood by an Inverse Wishart *a priori* for covariance matrices. In fact, we know that the origin of degeneracy is that the covariance matrices $\boldsymbol{R}_z$ and $\boldsymbol{R}_\epsilon$ approach the boundary of singularity (in a non arbitrary way). Thus, if we penalize the likelihood such that when one of the covariance matrices approaches the boundary, the *a posteriori* distribution goes to zero, eliminating the infinity value at the boundary and even forcing it to zero.

**Proposition** $5$**:** $\forall\ \boldsymbol{x}_{1..T} \in (\mathrm{R}^m)^T$, the likelihood $p(\boldsymbol{x}_{1..T}\,|\,\boldsymbol{\theta}_z, \boldsymbol{R}_\epsilon, \boldsymbol{A})$ penalized by an *a priori* Inverse Wishart for the noise covariance matrix $\boldsymbol{R}_\epsilon$ or by an *a priori* Inverse Wishart for the matrices $\boldsymbol{R}_z$ is bounded and goes to $0$ when one of the covariance matrices approaches the boundary of singularity.

**Proof 5:** The proof is based upon the proof of the proposition $4$, except the fact that here the *a priori* is not directly related to the matrices $\boldsymbol{\Gamma}_z = \boldsymbol{A}\boldsymbol{R}_z\boldsymbol{A}^* + \boldsymbol{R}_\epsilon$, but to covariance matrices $\boldsymbol{R}_z$ or $\boldsymbol{R}_\epsilon$. Then, we have the following alternative:

- If one penalizes by an *a priori* Inverse Wishart on the matrix $\boldsymbol{R}_\epsilon$, we have the following inequality:

$$(p(\boldsymbol{R}_\epsilon))^{1/T}\mathcal{N}(\boldsymbol{A}\boldsymbol{\mu}_z, \boldsymbol{\Gamma}_z) \leq \frac{A}{|\boldsymbol{\Gamma}_z|^{1/2}}\frac{K}{|\boldsymbol{R}_\epsilon|^{\beta}}\exp\left[-\alpha\,\mathrm{Tr}\left(\boldsymbol{R}_\epsilon{}^{-1}\boldsymbol{J}\right)\right]$$

  Now according to the mini-max principle applied to the characterization of eigen-values, we have:

$$|\boldsymbol{\Gamma}_z| = |\boldsymbol{A}\boldsymbol{R}_z\boldsymbol{A}^* + \boldsymbol{R}_\epsilon| \geq |\boldsymbol{R}_\epsilon|$$

which yields the following inequality:

$$(p(\boldsymbol{R}_\epsilon))^{1/T}\mathcal{N}(\boldsymbol{A}\boldsymbol{\mu}_z,\boldsymbol{\Gamma}_z) \leq \frac{1}{|\boldsymbol{R}_\epsilon|^b}\exp\left[-a\,\mathrm{Tr}\left(\boldsymbol{R}_\epsilon^{-1}\,\boldsymbol{J}\right)\right]$$

This insures the convergence to $0$ of the penalized likelihood when $\boldsymbol{R}_\epsilon$ goes to a singular matrix and insures, as well, the elimination of degeneracy which one the necessary conditions is the singularity of the covariance $\boldsymbol{R}_\epsilon$.

- If we penalize only by an Inverse Wishart prior on the matrices $\boldsymbol{R}_z$ with an uniform *a priori* on the matrix $\boldsymbol{R}_\epsilon$, we have a similar inequality:

$$(p(\boldsymbol{R}_z))^{1/T}\mathcal{N}(\boldsymbol{A}\boldsymbol{\mu}_z,\boldsymbol{\Gamma}_z) \leq \frac{1}{|\boldsymbol{A}\boldsymbol{R}_z\boldsymbol{A}^*|^{b_z}}\exp\left[-a_z\,trace\left(\boldsymbol{R}_z^{-1}\,\boldsymbol{J}_z\right)\right]$$

Here, the only query is that the determinant $|\boldsymbol{A}|$ goes to zero faster than the exponential of $|\boldsymbol{R}_z|$ but, in this situation, the degeneracy condition (3) is not verified because of the inclusion relation (4).

## CONCLUSION

The set of parameter singularities which characterizes the likelihood degeneracy of a multivariate Gaussian mixture is identified. A Bayesian solution to this degeneracy is proposed. We proposed a modified version of the data augmentation algorithms which allows to account for some constraints on the structure of the covariance matrices of the Gaussian mixture distribution. It consists essentially in the introduction of an inverse iteration to make the problem linear with respect to the matrix estimate. The case of source separation with Gaussian mixture model sources is also considered and discussed.